\documentclass[conference]{IEEEtran}

\usepackage[T1]{fontenc}

\usepackage{amssymb}	 % For \union.
\usepackage{amsmath}

\usepackage{graphicx}
\usepackage{booktabs}
\usepackage{multirow}
\usepackage{tabularx}
\usepackage{enumitem}
\usepackage{threeparttable}

\newcommand{\vc}[1]{\boldsymbol{#1}}

\newcommand{\mtp}{\textup{\textbf{\textrm{P}}}}

\newcommand{\vcdelta}{\vc{\delta}}

\newcommand{\vcphi}{\vc{\phi}}
\DeclareMathOperator{\I}{I}

\usepackage{balance}

\usepackage{url}
\binoppenalty=\maxdimen
\relpenalty=\maxdimen

\begin{document}

\title{How PHP Releases Are Adopted in the Wild?}
\author{
\IEEEauthorblockN{Jukka Ruohonen}
\IEEEauthorblockA{University of Turku, Finland \\ 
Email: juanruo@utu.fi}
\and
\IEEEauthorblockN{Ville Lepp\"anen}
\IEEEauthorblockA{University of Turku, Finland \\ 
Email: ville.leppanen@utu.fi}
}

\maketitle

\begin{abstract}
This empirical paper examines the adoption of PHP releases in the the contemporary world wide web. Motivated by continuous software engineering practices and software traceability improvements for release engineering, the empirical analysis is based on big data collected by web crawling. According to the empirical results based on discrete time-homogeneous Markov chain (DTMC) analysis, (i)~adoption of PHP releases has been relatively uniform across the domains observed, (ii) which tend to also adopt either old or new PHP releases relatively infrequently. Although there are outliers, (iii) downgrading of PHP releases is generally rare. To some extent, (iv)  the results vary between the recent history from 2016 to early 2017 and the long-run evolution in the 2010s. In addition to these empirical results, the paper contributes to the software evolution and release engineering research traditions by elaborating the applied use of DTMCs for systematic empirical tracing of online software deployments.
\end{abstract}


\section{Introduction}

Programming languages evolve like any other software~\cite{Favre05}. Like most software, also programming languages require release engineering, and as with conventional software, users of a programming language are likely to abandon the language if it is not properly updated and maintained to meet the continuously changing requirements \cite{Meyerovich12}. In recent years, different continuous software engineering practices have become increasingly popular for the development and maintenance of conventional software. Interestingly, also programming languages such as PHP have adopted a strategy of continuous releases scheduled to occur in a fast and fixed release cycle.

This paper investigates the continuous release engineering of the PHP programming language from a perspective of deployments using the language to serve some of the most popular web sites in the current Internet. While the release engineering practices used by the PHP project establish the practical motivation, the primary scholarly purpose of the investigation is to examine the previously unexplored use of classical DTMCs for studying release engineering. For putting the elaborated DTMCs into work, large web crawling datasets are used to analyze the current PHP release adoption patterns. 

Markov chains belong to the classical methodology toolbox in reliability engineering \cite{Cheung80, Wang12}. Discrete time-homogeneous Markov chains have recently been also adopted for studying different empirical software engineering problems, including those related to software evolution \cite{WongCai11}. Thus far, however, empirical applications have been limited in the release engineering domain. In addition to patching this limitation in the literature, this paper contributes to the release-based approaches (as opposed to approaches based on version control and bug tracking systems)  for studying software evolution.

There exists a large amount of empirical work using a release-oriented perspective to study software evolution in general and release engineering in particular. Backward-compatibility~\cite{Raemaekers14}, library dependencies \cite{Kula15}, and so-called rapid releases~\cite{Karvonen17, Mantyla13, daCosta16} are good examples of recent questions examined  (for a review of current research challenges see~\cite{Adams16}). Many of these studies deal with upgrading and downgrading question either explicitly or implicitly. There is thus plenty of prior work for framing the questions examined. 

However, much of the existing empirical work has literally been release-based, whereas this paper leans towards a deployment-based approach. In other words, much of the prior work is on the producer-side, while the consumer-side has received less attention \cite{Baysal12}. While bringing these two sides closer to each other remains a major research challenge, it is worthwhile to further note that the release-versus-deployment distinction applies also to studies examining the evolution of programming languages. In particular, there is some prior work examining PHP deployments seen in the wild~\cite{Ruohonen16WIMS}, but the evolution of the programming language itself---with its features and flaws---has received more attention \cite{Amanatidis16, Hills15}. The same applies to PHP source code analysis, which has mostly concentrated on evaluating ``off-the-shelf'' PHP applications (e.g., \cite{Hills13, Medeiros16}) without attempting to cover custom applications seen in the wild. Although source code analysis is not pursued in this paper, the deployments examined still cover many custom PHP applications. Due to such applications, better knowledge about adoption and patching on the consumer-side is valuable for those on the producer-side.
In other words, it is important to consider continuous tracing of software deployments in order to improve the feedback loops required for sound continuous software engineering practices.

The remainder of the paper is structured into four sections. Section~\ref{section: background} motivates the research background in more detail and formulates the research questions for the empirical analysis. Section~\ref{section: approach} outlines the DTMC modeling approach. Results are presented in Section~\ref{section: experimental results} based on large longitudinal datasets compiled from a few third-party web crawling snapshots. Conclusions and discussion follow in Section~\ref{section: discussion}.

\section{Background}\label{section: background}

The scholarly background can be motivated by considering the feedback channels that are essential for the contemporary continuous software engineering practices. After connecting these practices to the concept of software traceability, research questions are formulated in relation to the current release engineering strategy of the PHP programming language.

\subsection{Motivation}

Continuous software engineering is an umbrella term covering multiple contemporary software engineering tools and methodologies, including but not limited to continuous planning, continuous budgeting, continuous integration, continuous delivery, continuous deployment, continuous testing, continuous evolution, continuous maintenance, continuous feedback, and, ultimately, continuous innovation~\text{\cite{Fitzgerald15, PangHindle16}}. These overlapping continuous-prefixed concepts are also well-recognized in the release engineering research domain.

However, much of the existing research has concentrated on traditional software engineering aspects, such as integration, build systems, testing, and maintenance. This emphasis is reflected in the attempts to define the concept of release engineering. For instance, release engineering has been defined as ``a  software engineering discipline  concerned  with the  development,  implementation, and improvement of processes to deploy high-quality software reliably and predictably'' \cite{Dyck15}. Although the word improvement appears in the definition, a little emphasis is placed on feedback from customers, investors, and other stakeholders, which is a fundamental element in the contemporary continuous software engineering practices and processes~\text{\cite{Karvonen17, Leppanen15}}. By and large, these feedback mechanisms have constituted an enduring challenge for empirical software evolution research in general \cite{deOliveira16, Ruohonen15JSEP}. This gap in the literature is noteworthy because the availability of information about releases has increased substantially in recent years. 

\begin{figure}[th!b]
\centering
\includegraphics[width=\linewidth, height=3.5cm]{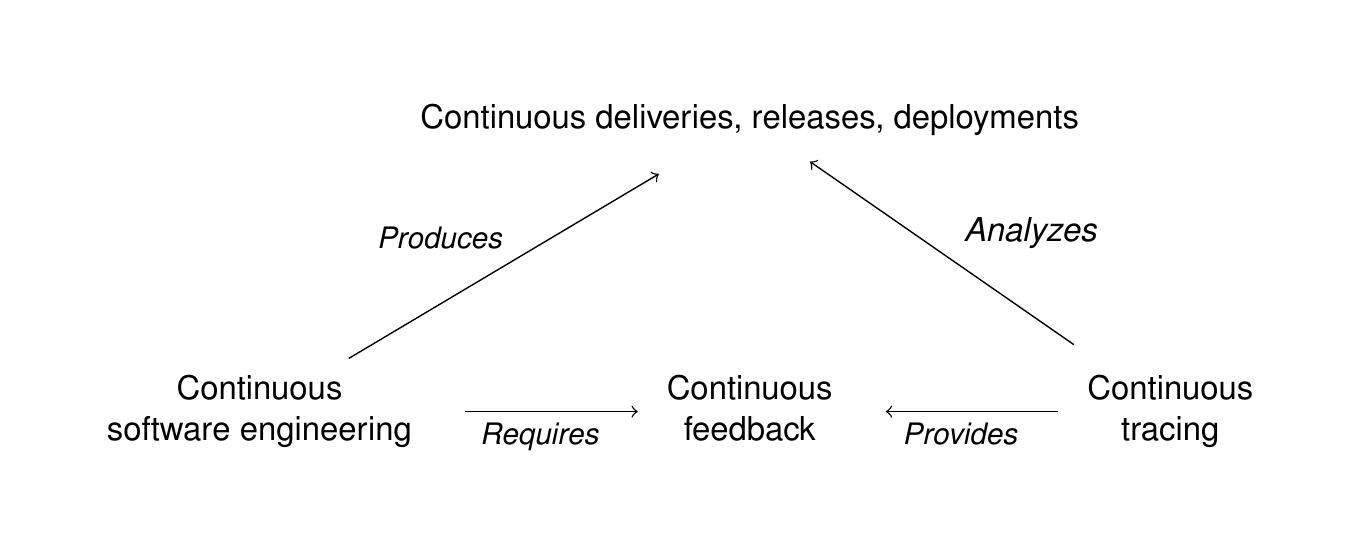}
\caption{A Terminological Map}
\label{fig: map}
\end{figure}

Different ``telemetry'' solutions---including crash reports and other ``call-home'' features---are increasingly popular in many software industry segments. The availability of feedback data is not limited to features explicitly integrated into software, however. Social media, review and rating sites, and related elements of the contemporary world wide web provide a wealth of information for systematic tracing of releases. A~major challenge for modern release engineering relates to integration of such data into meaningful solutions that help developers and stakeholders to make informed decisions about the evolution and patching of software deployments \cite{Adams16}. For summarizing this key challenge, Fig.~\ref{fig: map} depicts a relational map of a few interconnected continuous software engineering concepts. In this paper, the focus is at the lower-right corner, which is labeled as continuous tracing of the continuously engineered releases that are continuously deployed in the wild.

The concept of continuous tracing can be linked to software traceability, which ``refers to the ability to describe and follow the life of a requirement'', release, or other software artifact ``in both a forwards and backwards direction'' \cite{Gotel94}. This definition can be used to clarify and frame the scope of this paper. 

By appending the word continuous to the traceability term, it is emphasized that tracing of software artifacts should be systematic and continuous throughout the life cycles of the artifacts. In this paper, the software engineering artifacts are releases of the PHP programming language, but the units of analysis are deployments using the releases for serving web pages. The tracing in forward and backward directions is done by observing upgrading (i.e., roll-forward), downgrading (i.e., roll-back or reverting), and release adoption (i.e., either upgrading or downgrading) patterns of PHP deployments.

\subsection{PHP Releases and Research Questions}\label{subsec: research questions}

The first version of the PHP programming language was announced in 1995. The second, third, fourth, and fifth major versions followed in 1997, 1998, 2000, 2004, respectively. The sixth major version was branched for development in 2010. Instead of evolving into a production-ready major release branch, the controversies regarding Unicode support resulted in backporting of features from PHP~6 to the fifth major branch~\cite{Sturgeon14}. Currently, most PHP deployments still run with PHP~5, while the head of development occurs in the PHP~7 branch, which is not compatible with the previous major branches due to numerous new language features. For the programming language developers involved in the project, it is relevant to know an answer to the following question (\ref{rq: php7}).

\begin{enumerate}[label=RQ$_\arabic{enumi}$]
\item{\textit{How widespread has the transition been to PHP~7?}}\label{rq: php7}
\end{enumerate}

The PHP project follows the semantic versioning strategy conveyed via the ``\textit{major.minor.maintenance}'' versioning scheme. According to this versioning strategy, in essence, a major release should be reserved for incompatible changes to the application programming interfaces (APIs); a minor release should aggregate functionality enhancements that are backward-compatible; and, finally, maintenance releases should be reserved for small backward-compatible bug fixes and reliability improvements~\cite{Raemaekers14, Kula15}. These versioning principles have also guided the PHP release process since 2010 when a fixed release cycle was agreed upon. According to the current strategy, minor releases are scheduled to occur annually, whereas maintenance versions are released at least once a month \cite{PHP10}. Backward-compatibility and API stability are guaranteed within major branches. At the time of writing, the 5.6, 7.0, and 7.1 minor branches are still supported \cite{PHP17b}, which is in accordance with the guarantee of three years of support (bug and security fixes) for each minor release.

\begin{figure*}[th!b]
\centering
\includegraphics[width=\linewidth, height=7.5cm]{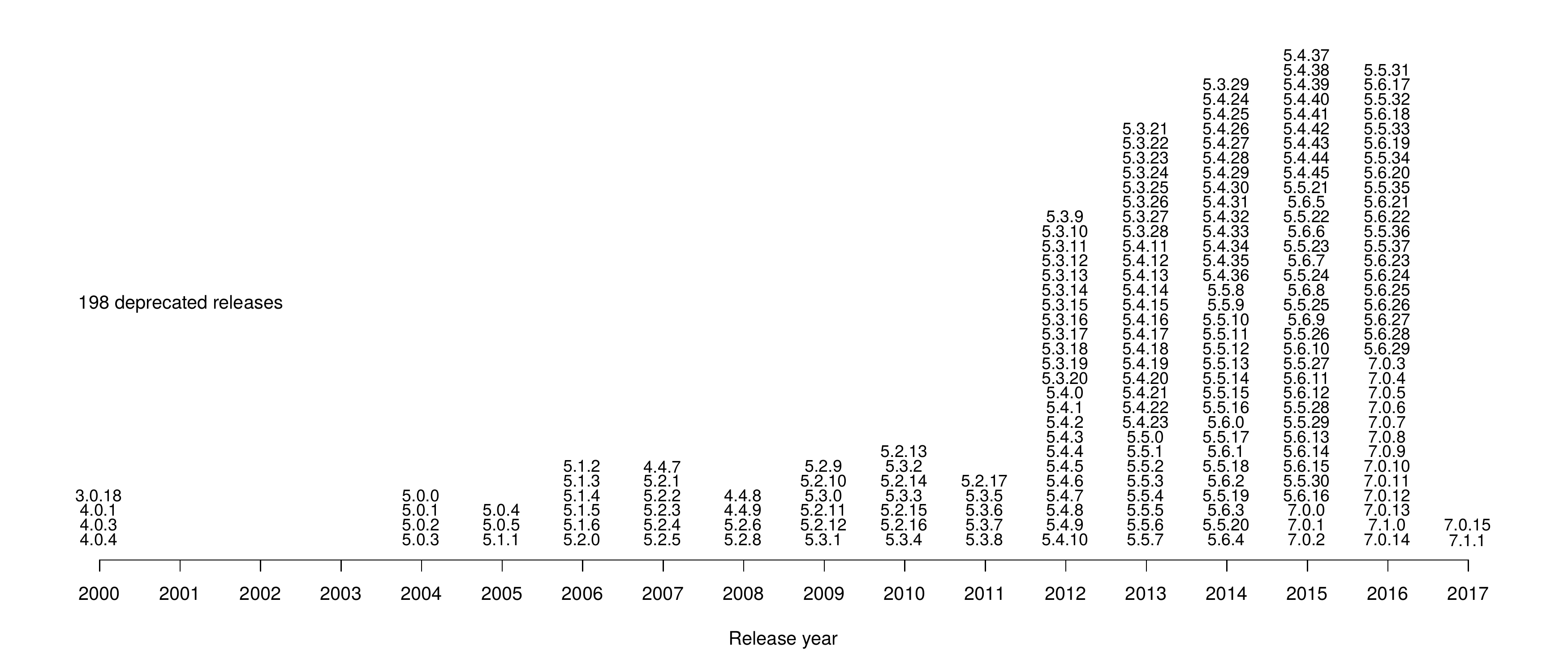}
\caption{Documented PHP Release History (unsupported versions as of March 2017, parsed from \cite{PHP17a})}
\label{fig: releases}
\end{figure*}

For a programming language project, the monthly cycle for maintenance releases is extremely rapid. On paper, this cycle is actually faster than those used for the development of many web browsers, such as Firefox \cite{Mantyla13}. As is visible from the illustration in Fig.~\ref{fig: releases}, the strategy of monthly maintenance releases has also resulted a large amount of versions from circa 2012 onward. Given the rapid release cycle and the large amount of releases made in recent years, it is relevant to solicit an answer to the research question \ref{rq: prevalence}. 

\begin{enumerate}[label=RQ$_\arabic{enumi}$, resume]
\item{\textit{How prevalent has the adoption of PHP releases been?}}\label{rq: prevalence}
\end{enumerate}

The question about release adoption includes both upgrading and downgrading of PHP versions. From a release engineering perspective, particularly interesting are cases whereby a deployment downgrades its PHP version. For instance: if a web site used a version \texttt{5.5.0} at some point in time but then later adopted a version \texttt{5.4.0}, perhaps there were difficulties in adopting the new release. If such downgrading is common, it might be worthwhile to revisit the supportive activities \cite{Mantyla11} associated with release engineering. Actual bug fixes notwithstanding, these activities include sufficient release notes, good and up-to-date documentation, user support, easy installation procedures, pre-install checks, sane defaults, migration instructions, and related release engineering aspects. Given this reasoning, the question (\ref{rq: downgrading}) is worth asking. 

\begin{enumerate}[label=RQ$_\arabic{enumi}$, resume]
\item{\textit{How common is downgrading of PHP deployments?}}\label{rq: downgrading}
\end{enumerate}

At a more abstract level of thought, it is relevant to know how consistent or uniform release adoption has generally been in recent history. By uniformity, it is meant that most deployments follow the semantic versioning strategy in their upgrades, moving within a major or minor branch in a relatively logical manner. When planning for new releases or supportive activities thereto, it is less relevant to try to support deployments that adopt releases in a chaotic manner. For instance: if a site upgraded from \texttt{5.4.0} to \texttt{7.1.1} but then moved to \texttt{5.5.0} while using a version \texttt{5.3.1} in-between, there are likely problems in the maintenance of the site, which cannot be addressed by the means of release engineering. If such chaotic patterns are widespread, on the other hand, it may be relevant to reconsider the appropriateness of a release strategy. Thus, the following question (\ref{rq: uniform}) is justified.

\begin{enumerate}[label=RQ$_\arabic{enumi}$, resume]
\item{\textit{How uniform has the adoption PHP releases been?}}\label{rq: uniform}
\end{enumerate}

Finally, the following atheoretical assertion can be placed for controlling the answers to the research questions outlined.

\begin{enumerate}[label=RQ$_\arabic{enumi}$, resume]
\item{\textit{Do the answer to \ref{rq: php7}, \ref{rq: prevalence}, \ref{rq: downgrading}, and \ref{rq: uniform}  vary between the recent short-run history and the long-run evolution?}}\label{rq: short-run vs. long-run}
\end{enumerate}

The concepts of prevalence, uniformity, short-run, and long-run are further elaborated in the subsequent sections that introduce the Markov chain framework and the empirical data.

\section{Approach}\label{section: approach}

A few remarks about the fundamental properties of DTMCs are required to outline the research approach. After these remarks, computation and operationalization are discussed.

\subsection{DTMCs in Brief}\label{subsec: dtmcs in brief}

%The assumptions from (a) to (c) are fundamental for a classical DTMC analysis (see \cite{Privault13} for a good introduction to the applied mathematical background). 

A first-order discrete time Markov chain is a finite sequence of random variables $X_1, X_2, \ldots, X_t, \ldots$, satisfying the fundamental Markov property according to which the probability distribution of a forthcoming $X_{t+1}$ depends on the immediately preceding $X_t$ but not on $X_{t-1}, X_{t-2}, \ldots, X_1$. If $S = \lbrace s_1, \ldots, s_n \rbrace$ denotes a set of all possible values of the random variables, the Markov property implies that the probability of moving to a next state in the state space $S$ is
\begin{align}\label{eq: Markov property}
\Pr&(X_{t+1} = s_{t+1} \vert~X_1 = s_1, X_2 = s_2, \ldots, X_t = s_t)
\\ \notag
&= \Pr(X_{t+1} = s_{t+1}~\vert~X_t = s_t) .
%\quad 0 < t \leq n - 1 .
\end{align}

This first-order Markov property implies a ``memoryless'' model, meaning that predicting a future state depends only on the current state. In addition to (a) assuming that~\eqref{eq: Markov property} holds, (b) the chains observed are assumed to be time-homogeneous. The latter condition means that a transition probability
\begin{equation}\label{eq: transition probability}
p_{ij} = \Pr(X_{t+1} = s_j ~\vert~ X_t = s_i)
\end{equation}
from a state $s_i \in S$ to state $s_j \in S$ is independent from $t$,
\begin{equation}
\Pr(X_{t+1} = s_j ~\vert~ X_t = s_i)
= \Pr(X_{t} = s_j ~\vert~ X_{t-1} = s_i) .
\end{equation}

In other words, the transition probabilities do not change as time passes. This assumption can be further accompanied by emphasizing that (c) only discrete chains are considered without explicit linkage to continuous calendar-time. This further restriction implies that the transition probability in \eqref{eq: transition probability} does not depend on the calendar-time lag between $s_j$ and $s_i$, irrespective whether the lag is measured in months or years.

Finally, (d)~the state changes associated with two distinct (exogenous) sequences, $X_1, X_2, \ldots, $ and $Y_1, Y_2, \ldots$, are assumed to be independent from each other, such that
\begin{align}\label{eq: exogenous}
\Pr(&X_{t+1} = s_j ~\vert~ X_t = s_i, Y_1 = s_1, \ldots, Y_t = s_k) ,
\\ \notag
&= \Pr(X_{t+1} = s_j ~\vert~ X_t = s_i) .
\end{align}
In other words, the cross-sectional empirical analysis is conducted without considering any potential dependencies between individual sequences and their state changes. 

\subsection{Computation}

The empirical setup is based on a sample of $m$ domains:
\begin{align}\label{eq: sequences}
X^{(1)}_1, &~X^{(1)}_2, \ldots, X^{(1)}_{r_1} 
\\ \notag
&\vdots
\\ \notag
X^{(m)}_1, &~ X^{(m)}_2, \ldots, X^{(m)}_{r_m} 
\end{align}
that are exogenous with respect to each other, such that \eqref{eq: exogenous} holds for any pair of sequences and their state changes. 

Due to practical reasons stemming from data collection, the length of the sequences and state changes are both allowed to vary across domains. For instance the length of the sequence for the $k$:th domain, denoted by $r_k$, may differ from another sequence length $r_{k+1}$. These varying sequence lengths correspond with the times each domain is observed empirically. 

As described later in Section~\ref{subsec: snapshots}, the maximum sequence lengths are 14 and 6 for all domains in the short-run and long-run examinations, respectively. In addition, a constraint $r_k \geq 2$ is imposed for all $m$ domains to ensure that state changes are possible to begin with. Even when the $k$:th domain is observed fourteen times, however, the length of the state space may equal one in case the domain in question never changed the PHP version of its deployment. In contrast, the maximum value $r_k$ for $\vert S_k \vert$ is attained by a domain that has changed its PHP version each time the domain is observed.

The transition probabilities are estimated by
\begin{equation}\label{eq: MLE}
\hat{p}^{(k)}_{ij} = 
\begin{cases}
0 & \textmd{if}~f^{(k)}_{i.} = 0, \\
f^{(k)}_{ij}~/~f^{(k)}_{i.} & \textmd{if}~f^{(k)}_{i.} \neq 0 ,
\end{cases}
\end{equation}
where $f^{(k)}_{ij}$ denotes the frequency of $(X_t = s_i, X_{t+1} = s_j)$ PHP version sequences for the $k$:th domain and 
\begin{equation}
f^{(k)}_{i.} = \sum^{\vert S_k \vert}_{j=1} f^{(k)}_{ij} .
\end{equation}

The special case $f^{(k)}_{i.} = 0$ occurs when the last observed state denotes a previously unseen PHP version, meaning that there is not enough data to estimate the transition probability for this state. This additional, context-specific alteration notwithstanding, the equation \eqref{eq: MLE} conveys a conventional maximum likelihood estimator (MLE) for a transition probability from the $i$:th to the $j$:th state \cite{Singer14, Hill04}. While a small custom implementation is used for the MLE computations, the results were further verified with an existing R implementation \cite{markovchain}. %The same applies for standard errors (SEs) for the estimates, which are readily available from
%
%\begin{equation}\label{eq: MLE SE}
%\SE_{\hat{p}^{(k)}_{ij}} = 
%\begin{cases}
%1 & \textmd{if}~f^{(k)}_{ij} = 0, \\
%\hat{p}^{(k)}_{ij}~/~\sqrt{f^{(k)}_{ij}}
%& \textmd{if}~f^{(k)}_{ij} > 0  .
%\end{cases} 
%\end{equation}

\begin{figure}[th!b]
\centering
\includegraphics[width=7cm, height=5cm]{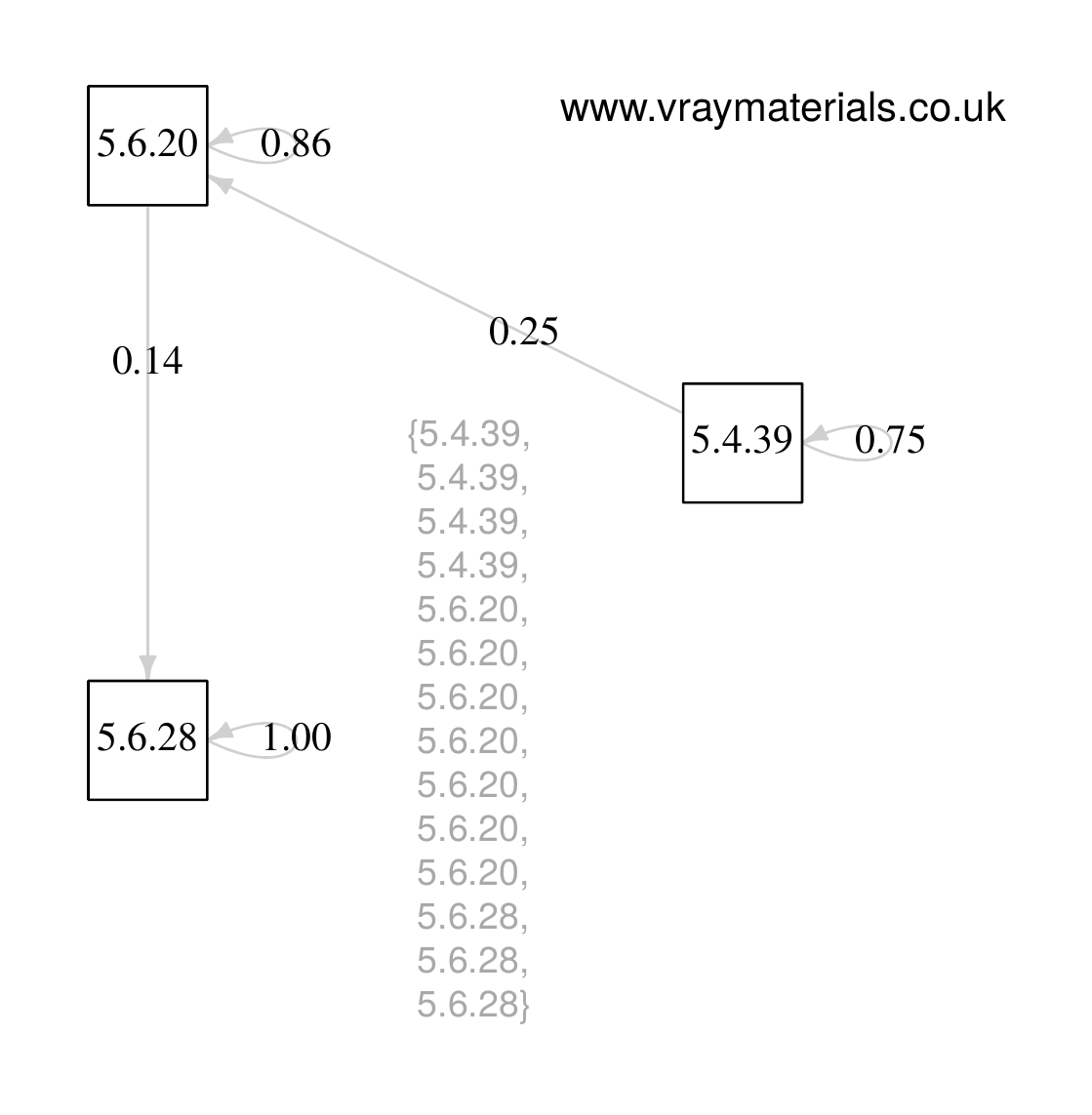}
\caption{An Example DTMC for PHP Release Adoption}
\label{fig: example 1}
\end{figure}

To illustrate the computation in practice, consider the example in Fig.~\ref{fig: example 1}. A full sequence of fourteen observations is available for this domain, while the state space contains three unique PHP versions. The first state corresponds with the PHP version \texttt{5.4.39}. Because the domain used the same deployment also during the three subsequent observations, the probability of upgrading from this version to \texttt{5.6.20} was $1/4 = 0.25$. The probability of subsequently upgrading the deployment from \texttt{5.6.20} to \texttt{5.6.28} is even lower, given the seven times the domain \texttt{www.vraymaterials.co.uk} stayed with its \texttt{5.6.20} deployment. Thus, the prevalence of release adoption has been modest for this particular domain in the short-run. For evaluating the prevalence among hundreds of thousands of web sites, a few custom metrics can be derived.

\subsection{Metrics}\label{subsec: metrics}

The research question about prevalence (\ref{rq: prevalence}) can be answered with a metric based on the estimated transition probabilities. Thus, let $\mtp_k$ denote a $\vert S_k \vert \times \vert S_k \vert$ matrix of estimated transition probabilities for the $k$:th domain. For instance, the $3 \times 3$ transition probability matrix underneath the illustration in Fig.~\ref{fig: example 1} is defined by
\begin{equation}
%\mtp_k \simeq
\underbrace{\left.\left[\begin{array}{ccc} 
0.75 & 0.25 & 0.00 \\
0.00 & 0.86 & 0.14 \\
0.00 & 0.00 & 1.00
\end{array}\right]\right\}}_{\texttt{5.4.39}~~\texttt{5.6.20}~~\texttt{5.6.28}~~}
\begin{array}{c}
\scriptstyle\texttt{5.4.39} , \\
\scriptstyle\texttt{5.6.20} , \\
\scriptstyle\texttt{5.6.28} , \\
\end{array} 
\end{equation}
which can be read as an adjacency matrix for a weighted and directed graph. The trace of this matrix (that is, the sum of the diagonal elements) provides a simple measure for the persistence of a DTMC phenomenon \cite{Hill04}. For answering to the first research question \ref{rq: prevalence}, this simple but powerful idea allows to operationalize the concept of prevalence with
\begin{equation}\label{eq: prevalence}
\delta_k 
= \frac{1}{\vert S_k \vert}\sum^{\vert S_k \vert}_{i=1} 
\left(1 - \hat{p}^{(k)}_{ii}\right),
\end{equation}
where $\delta_k \in [0, 1]$ for all $k$. In other words, the closer a $\delta_k$ is to unity, the more prevalent has the adoption of releases been. If $\delta_k = 0$, the $k$:th web site never changed its PHP deployment. Collecting the scalars to a vector $\vcdelta = [\delta_1, \delta_2, \ldots, \delta_m]$ allows evaluating the prevalence among the $m$ domains observed.

Answering to the research questions \ref{rq: downgrading} and \ref{rq: uniform} is better done with the version sequences in \eqref{eq: sequences} rather than with the transition probabilities within the state spaces. Thus, for evaluating how uniform PHP release adoption has generally been among the $m$ domains observed (\ref{rq: uniform}), a simple metric is available by counting the unique version sequences, scaling the resulting amount by $m$. Although this metric approaches zero as $m \to \infty$, it still gives a good overall sense about the uniformity of typical PHP release adoption patterns.

Although calendar-time records can be used for comparing release orderings \cite{Kula15}, a metric for downgrading (\ref{rq: downgrading}) can be also computed directly from the PHP version sequences. For all domains with $\vert S_k \vert \geq 2$, downgrading can occur via three different scenarios: (a) when $\textit{major}_{i+1} < \textit{major}_i$, that is, when the major version number of a current deployment is larger than the major version number of a subsequent deployment; (b) when  $\textit{major}_{i+1} = \textit{major}_{i}$ but $\textit{minor}_{i+1} < \textit{minor}_{i}$; or (c) when both the major and minor version numbers remain the same but the maintenance version number of the $i$:th state is larger than the number of the subsequent state. Given these three distinct cases, all $m$ version sequences are processed by comparing $(r_k - 1)$ times the $i$:th version to the $(i + 1)$:th version, recording the  number of downgrades at each step. If $d_k$ denotes the number of downgrades recorded for the $k$:th domain, a vector $\vcphi = [ d_1~/~r_1 - 1, \ldots, d_m ~/~ r_m - 1]$ defines a simple metric for evaluating how common PHP downgrading has generally been. Analogous to \eqref{eq: prevalence}, values close to unity indicate frequent downgrading. In theory, also different weights could be used for the three different downgrading scenarios, but this simple counting scheme is sufficient because downgrading should be relatively rare in the context of popular web sites.

The transition matrices $\mtp_1, \ldots, \mtp_m$ offer another viewpoint to downgrading: whenever states $s_i$ and $s_j$ communicate (such that there is a transition from $s_i$ to $s_j$ and from $s_j$ to $s_i$), there is also downgrading of PHP versions. Given this reasoning, a further metric can be computed by counting the number of communicating state pairs and scaling the result appropriately:
\begin{equation}\label{eq: communicating pairs}
\gamma_k = 
\frac{1}{r_k - 1}
\sum^{\vert S_k \vert}_{i=1}\sum^{\vert S_k \vert}_{j=i}
\I\left(\hat{p}^{(k)}_{ij}\right) 
\I\left(\hat{p}^{(k)}_{ji}\right) 
\end{equation}
where $\I(\cdot)$ is an indicator function outputting
\begin{equation}
\I(x) =
\begin{cases}
0 & \textmd{if}~x = 0, \\
1 & \textmd{if}~x > 0.
\end{cases}
\end{equation}

% "Larger than 3: b14.nakanohito.jp, idx = 137089, pairs = 4, l = 13, val = 0.307692307692308"
%
% and down.shortrun[137089] = 7.

\begin{figure}[th!b]
\centering
\includegraphics[width=7cm, height=5cm]{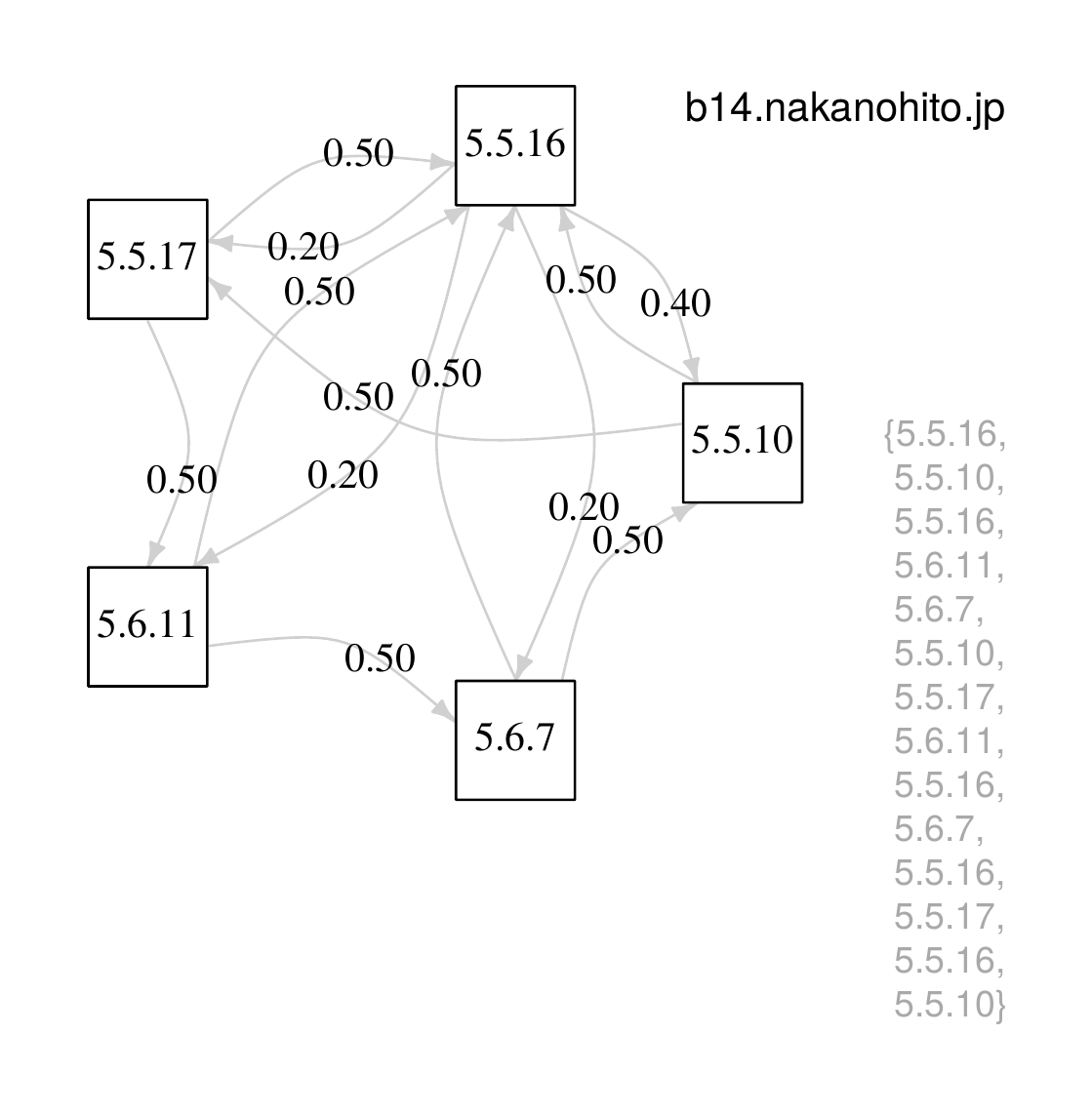}
\caption{Another Example DTMC for PHP Release Adoption}
\label{fig: example 2}
\end{figure}

As an example, consider the quite messy real-world release adoption pattern visualized in Fig.~\ref{fig: example 2}. This particular domain downgraded its PHP deployment as many as seven times during 2016 and early 2017. Therefore, $\phi_k = 7~/~(14 - 1) \simeq 0.54$ and $\gamma_k = 4~/~13 \simeq 0.31$, given the four communicating pairs. Both values are rather high, which indicates that the two downgrading metrics can be used also for probing outlying domains that may have problems with maintenance of their PHP deployments.

\section{Experimental Results}\label{section: experimental results}

The empirical results are disseminated by first introducing the data used for the DTMC computation. The metrics elaborated in the previous Section~\ref{subsec: metrics} are subsequently used for summarizing the empirical findings.

\subsection{Data}

Two datasets are used for the empirical analysis: one for observing short-run release adoption and the other for proxying the long-run evolution of PHP deployments seen in the wild. While calendar-time is not explicitly observed with the DTMCs computed, the definitions for short-run and long-run are still based on calendar-time: with one exception, the short-run dataset covers a period from January 2016 to March 2017 under a monthly sampling frequency, while the long-run dataset is based on annual records in a period between January 2012 and February 2017. Given the PHP release lineage illustrated in Fig.~\ref{fig: releases}, a truly long-run analysis should start already from the year 2000---or even earlier, but the historical periods used are imposed by the source of empirical data. This data source should be also elaborated in more detail.

\subsubsection{Snapshots}\label{subsec: snapshots}

Both datasets are compiled from a few large web crawling snapshots that contain data on hypertext transfer protocol (HTTP) headers used for identifying PHP versions. These open data snapshots are provided by the HTTP Archive web crawling project~\cite{HTTPArchive17a}, which has also been used in previous research \cite{Ruohonen16WIMS} alongside analogous archives \cite{Wambach16, Ainsworth13}. In total, fourteen crawling snapshots are used for compiling the short-run dataset (see Table~\ref{tab: short-run data}). Due to data availability issues, the long-run dataset is compiled only from six snapshots (see Table~\ref{tab: long-run data}), the earliest of which dates to January 2012. 

\begin{table}[th!b]
\centering
\begin{threeparttable}
\caption{Characteristics of the Short-Run Dataset \cite{HTTPArchive17a}$^a$}
\label{tab: short-run data}
\begin{tabular}{llccc}
\toprule
& Start of crawling & User-agent & Size (GB)$^b$ & PHP domains$^c$ \\
\cmidrule{2-5}
1. & March 1, 2017 & Chrome & $\simeq 47$ & 206,739 \\
2. & February 1, 2017 & Chrome & $\simeq 45$ & 202,772 \\
3. & December 2, 2016 & Chrome & $\simeq 51$ & 216,362 \\
4. & November 1, 2016 & Chrome & $\simeq 49$ & 220,871 \\
5. & October 1, 2016 & Chrome & $\simeq 51$ & 222,645 \\
7. & September 1, 2016 & Chrome & $\simeq 49$ & 222,051 \\
7. & August 1, 2016 & Chrome & $\simeq 48$ & 227,757 \\
8. & July 1, 2016 & Chrome & $\simeq 48$ & 225,623 \\
9. & June 1, 2016 & Chrome & $\simeq 56$ & 226,797 \\
10. & May 1, 2016 & Chrome & $\simeq 52$ & 217,592 \\
11. & April 1, 2016 & Chrome & $\simeq 50$ & 219,564 \\
12. & March 1, 2016 & Chrome & $\simeq 57$ & 230,028 \\
13. & February 1, 2016 & Chrome & $\simeq 52$ & 226,859 \\
14. & January 1, 2016 & Chrome & $\simeq 50$ & 225,362 \\
\bottomrule
\end{tabular}
\begin{scriptsize}
\begin{tablenotes}
\item[]{$^a$~Note that the January 1 snapshot from 2017 was empty and had to be thus excluded. Due to this omission, there is a two month calendar-time delay between the second and the third snapshot. $^b$ The size refers to the unpacked snapshots. $^c$~See~Section~\ref{subsec: pre-processing} for a definition of a ``PHP domain''.} 
\end{tablenotes}
\end{scriptsize}
\end{threeparttable}
\end{table}

\begin{table}[th!b]
\centering
\begin{threeparttable}
\caption{Characteristics of the Long-Run Dataset \cite{HTTPArchive17a}$^a$}
\label{tab: long-run data}
\begin{tabular}{llccc}
\toprule
& Start of crawling & User-agent & Size (GB)$^b$ & PHP domains$^c$ \\
\cmidrule{2-5}
1. & February 1, 2017 & Chrome & $\simeq 45$ & 202,772 \\
2. & January 1, 2016 & Chrome & $\simeq 50$ & 225,362 \\
3. & January 1, 2015 & IE$^d$ & $\simeq 41$ & 247,568 \\
4. & January 1, 2014 & IE$^d$ & $\simeq 24$ & 158,775 \\
5. & January 1, 2013 & IE$^d$ & $\simeq 22$ & 165,741 \\
6. & January 1, 2012 & IE$^d$ & $\simeq 3.7$ & ~31,445 \\
\bottomrule
\end{tabular}
\begin{scriptsize}
\begin{tablenotes}
\item[]{$^{a,b,c}$~See the notes in Table~\ref{tab: short-run data}. $^d$ The abbreviation stands for Internet Explorer.} 
\end{tablenotes}
\end{scriptsize}
\end{threeparttable}
\end{table}

As shown in the two tables, the raw snapshots used are quite large. Because the crawls are seeded from Alexa's list of top-million busiest web sites \cite{HTTPArchive17b}, which is updated daily, the snapshot sizes also vary from a crawl to another. Moreover, it should be emphasized that the dates shown are only tentative regarding individual HTTP requests and responses: due to the large seeding list, crawling can take a relatively long amount of time \cite{Ainsworth13}. Already because calendar-time is not explicitly observed, the issue is a minor concern for this paper, however. 

The long-run dataset is affected by a change in the forged user-agent \cite{Calzarossa14} used for making the requests. Although user-agents can have a substantial empirical effect for measuring web sites due to specific responses for specific browsers~\cite{Pham16}, the consequences should be small in this paper because it seems unlikely that PHP version strings in the HTTP response headers would vary according to a user-agent specified in the HTTP request headers. Therefore, it is more important to further remark that the long-run dataset is affected by changes made to the seeding of the web crawls, which is reflected in the smaller amount of PHP domains between 2012 and 2014. In contrast, each snapshot in the short-run dataset contains roughly the same amount of domains. Finally, it should be emphasized that the total amount of PHP-powered domains observed is substantially larger than reported in Tables~\ref{tab: short-run data}~and~\ref{tab: long-run data} because the snapshots are ``pooled'' to include all domains that are present in at least two snapshots.

\subsubsection{Pre-processing}\label{subsec: pre-processing}

The snapshots were pre-processed from the packaged archives delivered as CSV (comma-seperated value) files. Although the files are provided as open data, a couple of remarks should be made to ensure replicability of the datasets. First and foremost, the presence of PHP is identified via s simple (Python) regular expression of the following form: ``\verb|PHP/[0-9]{1}\.[0-9]{1}\.[0-9]{1,}|'', where the quotation marks are not part of the expression. Notice that the expression excludes ``invalid'' versions such as \texttt{PHP/3.100}. 

Second, unique domains are identified by extracting the network location from the uniform resource locators (URLs) crawled. Because multiple web pages may be crawled for each domain, duplicates are excluded by omitting the parsing of URLs for domains that have already been identified to run with PHP. It should be remarked that the concept of domain is inexplicit in the sense that no attempts are made to lookup the domains via the domain name system. Therefore, in theory, the domains may refer to actual domain names as well as Internet protocol addresses. For the purposes of this paper, the distinction is irrelevant, however.

\subsection{Results}

The dissemination of the results can be started by noting a few characteristics of the two datasets compiled from the web crawling snapshots. First and foremost, according to the numbers shown in Table~\ref{tab: lengths}, about 451 and 220 hundred thousand domains were identified as running with PHP according to the simple pre-processing routines. By implication, well over half a million transition probabilities were estimated via \eqref{eq: MLE}. Second, on average, about a half of the maximum lengths of the version sequences are realized in the two datasets, although the standard deviations are large. In other words, a typical domain is observed a little over six times in the short-run and about three times in the long-run. The reason for not reaching the maximum lengths is simple: because the snapshots are not crawled from a fixed domain set, not all of the PHP domains observed are present across all snapshots. Third, the average size of the state space, $\frac{1}{m}\sum^m_{k=1} \vert S_k \vert$, is less than two in both datasets. Thus, for many domains, the transition probabilities are represented by the value one supplied via a $1 \times 1$ matrix. While there is still a sufficient amount of variance for analysis, already this observation allows to conclude that the prevalence of PHP release adoption has been at a modest level. Before continuing to the actual prevalence metric, a remark should be made about the most common PHP releases in the datasets.

\begin{table}[th!b]
\centering
\begin{threeparttable}
\caption{Sample Characteristics}
\label{tab: lengths}
\begin{tabular}{llrr}
\toprule
&& Short-run & Long-run \\
\cmidrule{2-4}
Number of domains ($m$) && 451,340 & 220,293 \\
\cmidrule{2-4}
Number of versions ($r$) & Mean & 6.5 & 2.9 \\
& Std.~dev. & 4.0 & 1.1 \\
\cmidrule{2-4}
Size of state space ($\vert S \vert$) & Mean & 1.5 & 1.8 \\
& Std.~dev. & 1.2 & 0.8 \\
\bottomrule
\end{tabular}
%\begin{scriptsize}
%\begin{tablenotes}
%\item[]{} 
%\end{tablenotes}
%\end{scriptsize}
\end{threeparttable}
\end{table}

\begin{table}[th!b]
\centering
\begin{threeparttable}
\caption{Uniform PHP Version Sequences}
\label{tab: uniform sequences}
\begin{tabular}{llrr}
\toprule
& Subset & Short-run & Long-run \\
\cmidrule{2-4}
Number of unique sequences & $\vert S_k \vert = 1$ & 2,470 & 568 \\
& $\vert S_k \vert > 1$ & 62,376 & 43,585 \\
\cmidrule{2-4}
Share of unique sequences (\%) & $\vert S_k \vert = 1$ & 0.55 & 0.26 \\
& $\vert S_k \vert > 1$ & 13.8 & 19.8 \\
\bottomrule
\end{tabular}
%\begin{scriptsize}
%\begin{tablenotes}
%\item[]{} 
%\end{tablenotes}
%\end{scriptsize}
\end{threeparttable}
\end{table}

\begin{figure}[th!b]
\centering
\includegraphics[width=\linewidth, height=4cm]{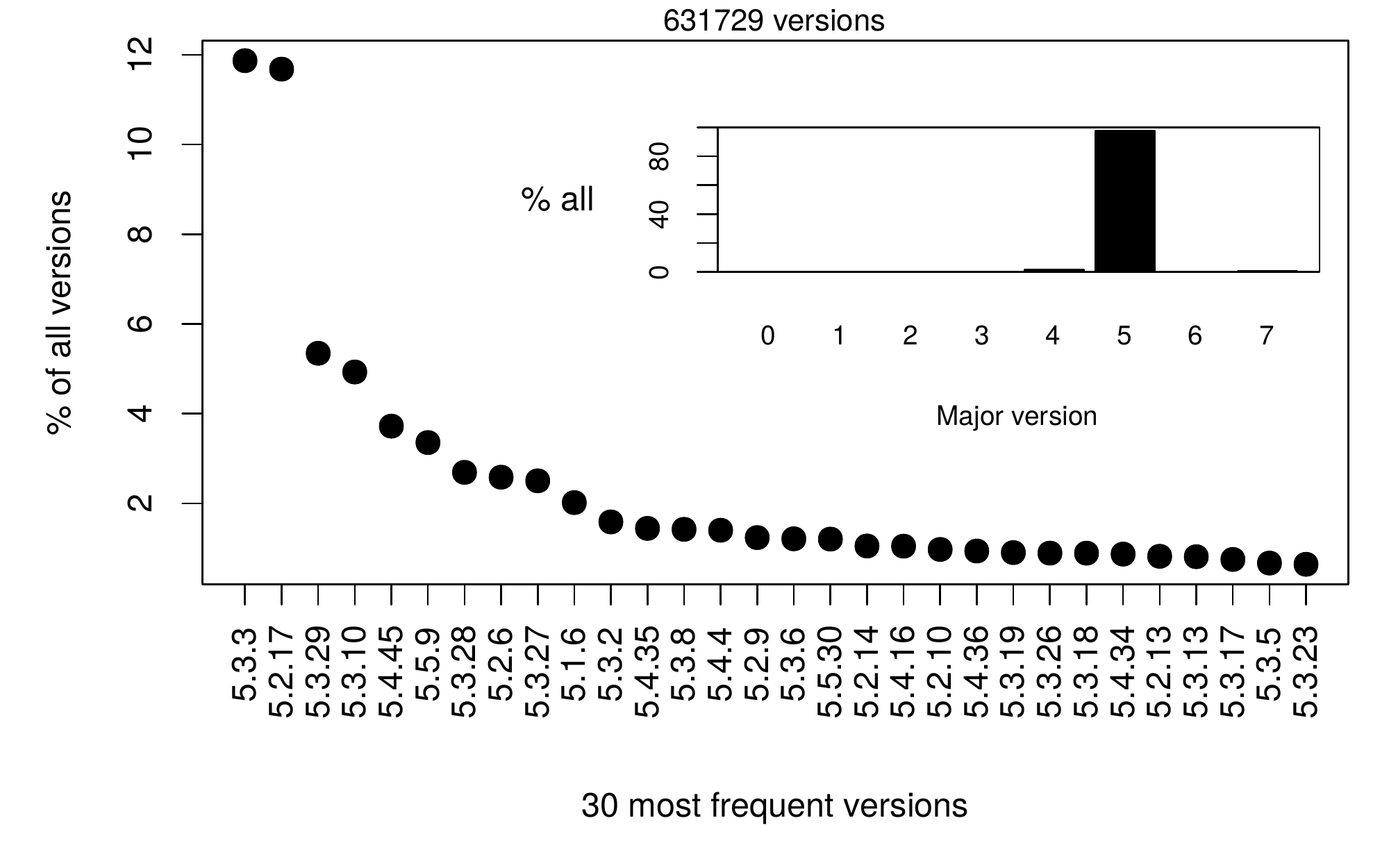}
\caption{Most Frequent PHP Releases in the Long-Run}
\label{fig: long-run releases}
\end{figure}

\begin{figure}[th!b]
\centering
\includegraphics[width=\linewidth, height=4cm]{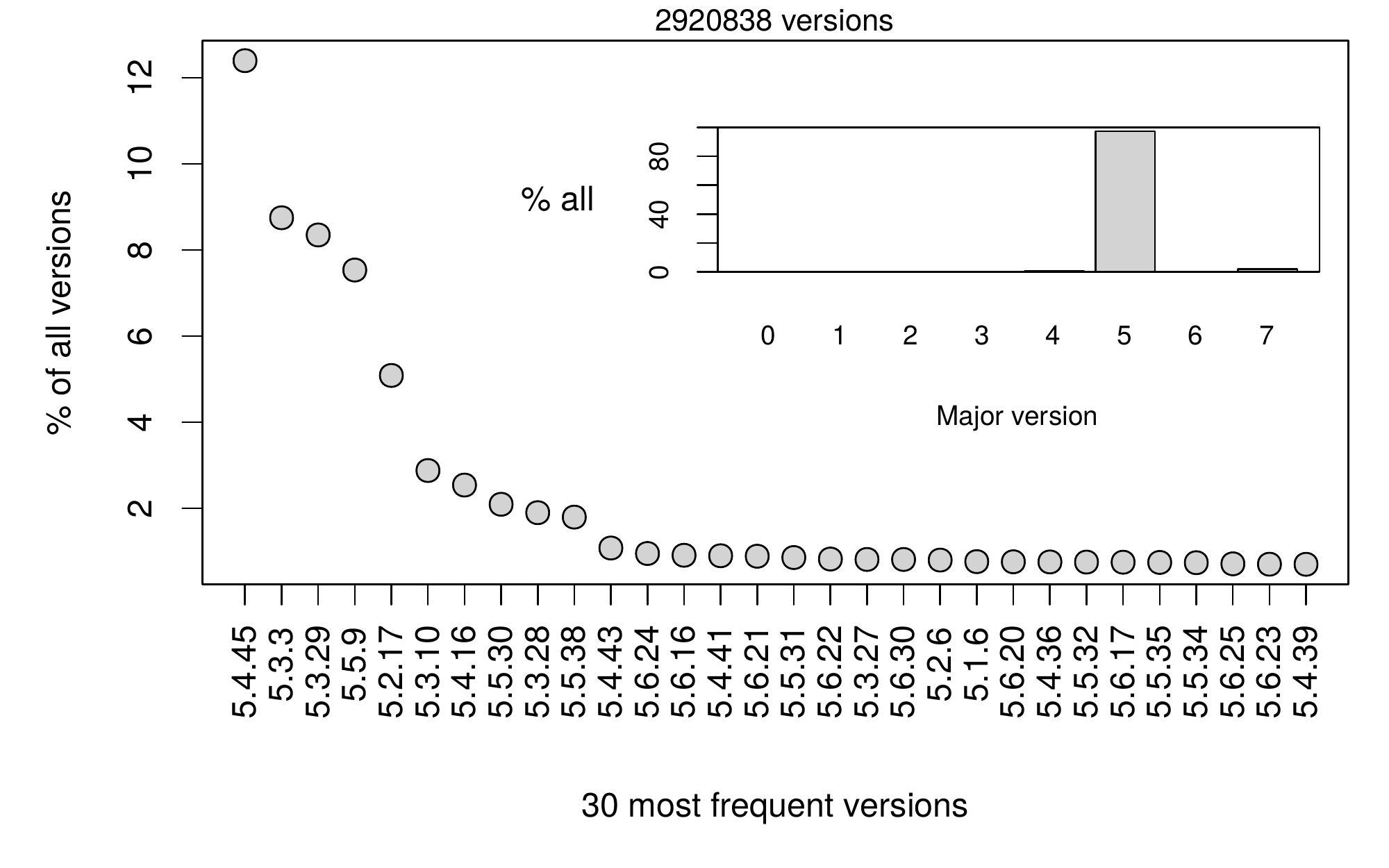}
\caption{Most Frequent PHP Releases in the Short-Run}
\label{fig: short-run releases}
\end{figure}

\begin{figure}[th!b]
\centering
\includegraphics[width=\linewidth, height=6cm]{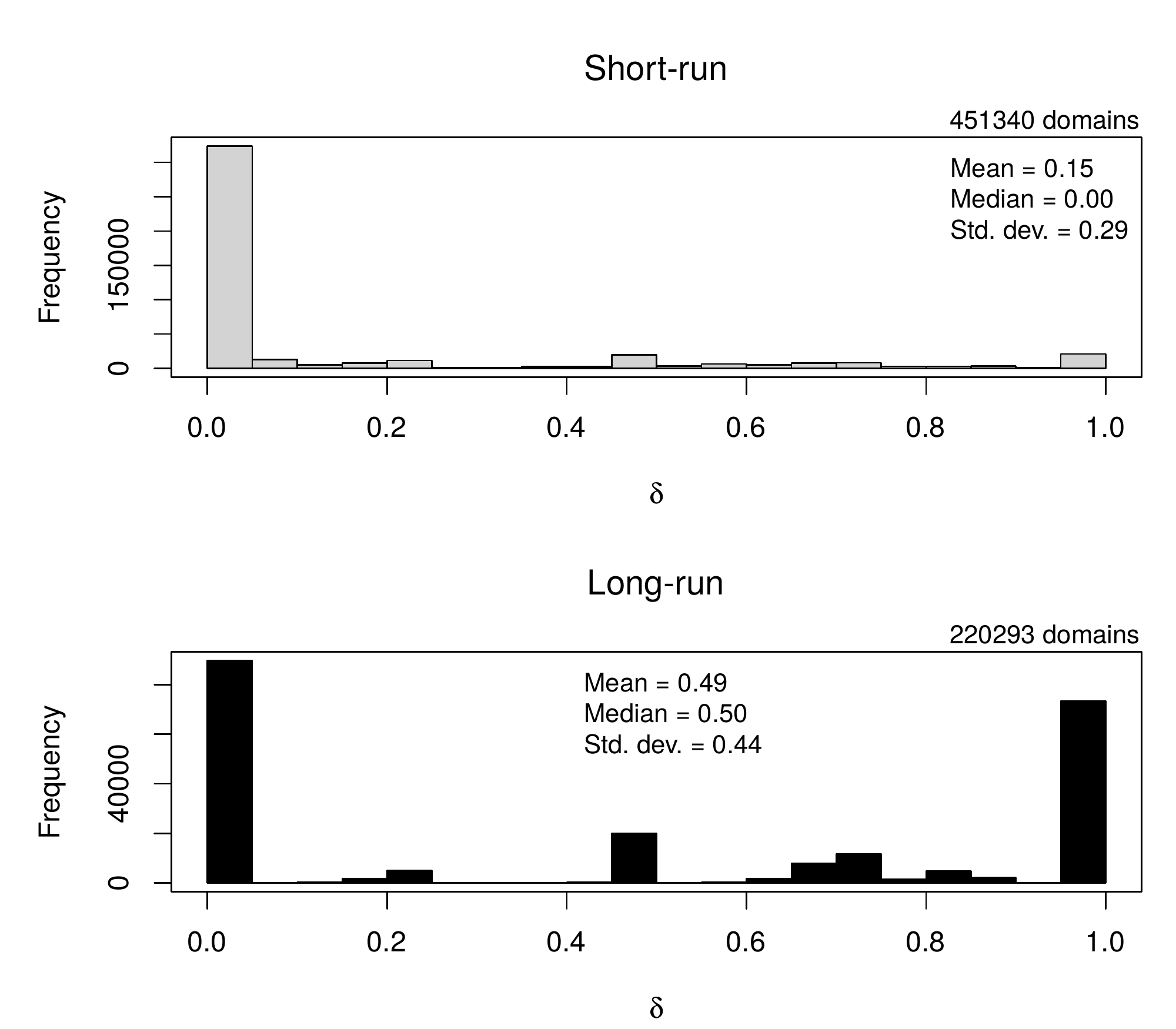}
\caption{Prevalence of PHP Release Adoption}
\label{fig: prevalence}
\end{figure}

\begin{figure}[th!b]
\centering
\includegraphics[width=\linewidth, height=6cm]{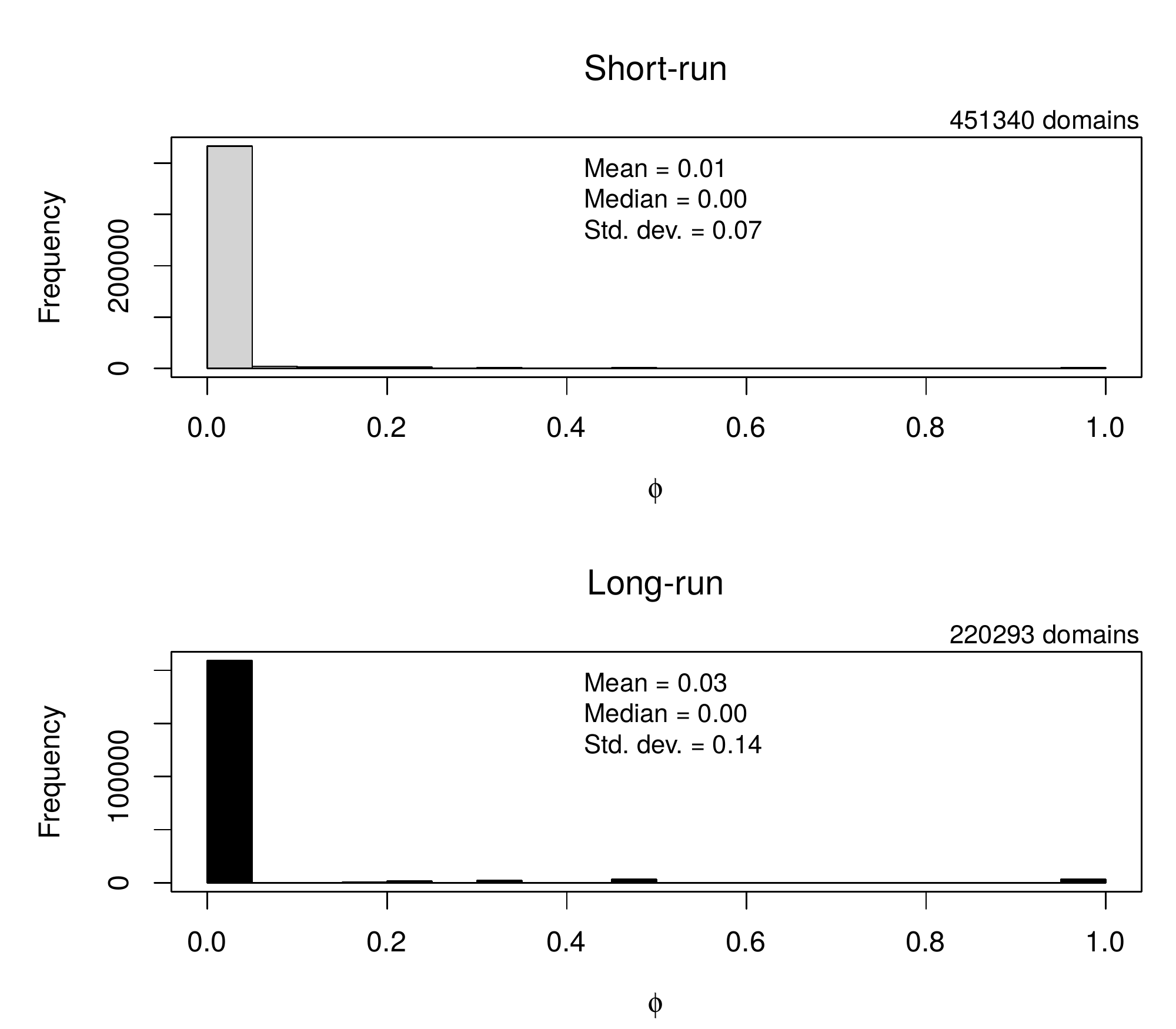}
\caption{Downgrading of PHP Releases \#1}
\label{fig: downgrading}
\end{figure}

\begin{figure}[th!b]
\centering
\includegraphics[width=\linewidth, height=6cm]{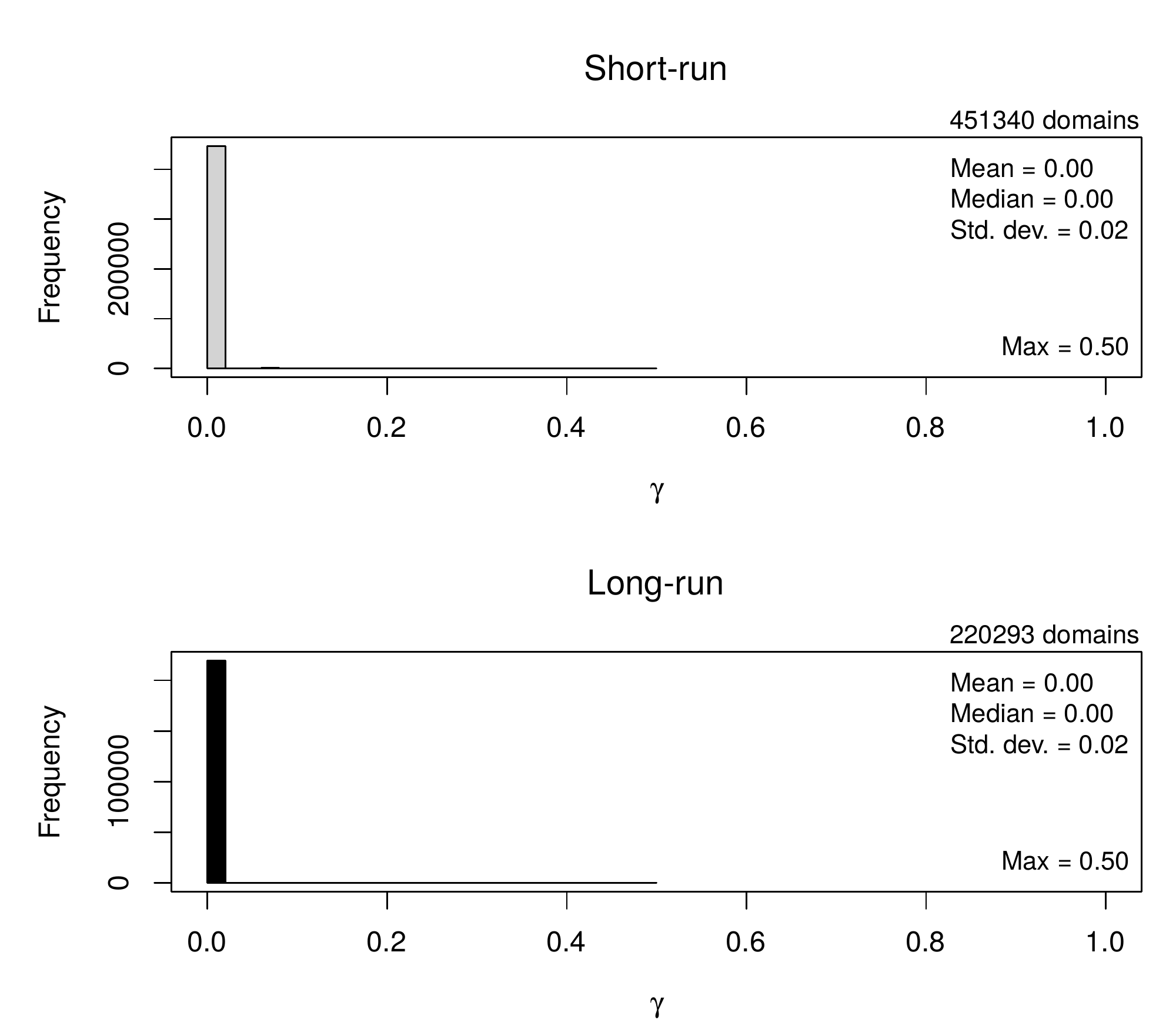}
\caption{Downgrading of PHP Releases \#2}
\label{fig: communicating}
\end{figure}

According to the datasets, it is clear that the PHP~5 release branch has been the most popular deployment choice from the early 2000s onward. As shown in Fig.~\ref{fig: long-run releases}, the clear majority of the versions observed in the long-run are part of the PHP~5 branch. While there has also been few domains using the PHP~4 branch, the use of other major branches has been negligible in the long-run. The fifth major version has also retained its popularity in 2016 and 2017, as can be concluded from the subsequent Fig.~\ref{fig: short-run releases}. Adoption of PHP~7 has been modest (\ref{rq: php7}). This conclusion does not change when only the last versions in the sequences are used to proxy popularity; in~this case, only about 1.7~\% of the domains have used a PHP~7 deployment in the short-run

The observation about relatively infrequent release adoption is reinforced by Fig.~\ref{fig: prevalence}, which shows a histogram of the metric in \eqref{eq: prevalence} across the domains observed in the two datasets. There exists a difference between short-run release adoption and long-run evolution, however. The upper plot clearly indicates that monthly release adoption has generally been infrequent in 2016 and early 2017. In contrast, the lower plot displays a bimodal distribution: many domains observed have not adopted releases in the long-run, but almost an equal amount of domains have adopted releases annually. All in all, the prevalence of PHP release adoption has been modest in recent years (\ref{rq: prevalence}), although less so in the long-run (\ref{rq: short-run vs. long-run}). 

Moreover, PHP release adoption patterns have been relatively uniform across the domains observed (\ref{rq: uniform}), as can be concluded from the summary shown in Table~\ref{tab: uniform sequences}. In total, only about $0.55 + 13.8 \simeq 14.4$ and $20.1$ percent of the observed PHP version sequences are unique in the short-run and long-run datasets, respectively. While it should be kept in mind that these relative amounts are affected by the large amount of domains observed, these relative amounts still indicate a modest amount of unique release adoption patterns. Most of the patterns in both datasets describe common transition paths within the PHP~5 major release branch. From a release engineering perspective, this observation is a positive finding: most domains follow other domains in their upgrading patterns. Disorderly release adoption patterns are relatively rare.

The observations about infrequent release adoption and uniformity are reinforced by Fig.~\ref{fig: downgrading}, which shows the frequency of the first downgrading metric in the two datasets. Most domains have not downgraded their PHP deployments even once. Although downgrading is slightly more common in the long-run, the standard deviations are generally small. By implication, the same observation applies also for the results regarding the second downgrading metric in \eqref{eq: communicating pairs}. As can be concluded from Fig.~\ref{fig: communicating}, only a very few of the version sequences involve communicating PHP version pairs. The averages and standard deviations are both negligible. To summarize, downgrading has been rare (\ref{rq: downgrading}), and there are no notable differences between short-run and long-run (\ref{rq: short-run vs. long-run}).

\section{Discussion}\label{section: discussion}

The remainder of this paper first summarizes the main empirical findings, then enumerates a few threats to validity, and finally concludes with a couple of new research directions.

\subsection{Summary of Results}

This empirical paper observed PHP release adoption in two datasets covering over a half a million Internet domains and three million PHP versions deployed within these domains. The main findings can be summarized by briefly answering to the five research questions outlined in Section~\ref{subsec: research questions}.

\begin{itemize}
\item{Adoption of PHP~7 has been modest (\ref{rq: php7}). As of early 2017, only few popular web sites have adopted the new major release branch. Most sites continue to operate with releases made within the PHP~5 major branch.}
\item{The prevalence of PHP release adoption has been at a modest level: popular web sites tend to upgrade their deployments relatively infrequently (\ref{rq: prevalence}). The observation aligns with previous studies; relatively old PHP~5 versions are commonly used in cloud computing services \cite{WangNappa14}.}
\item{Downgrading has been uncommon; only a few outlying domains have downgraded their deployments (\ref{rq: downgrading}).}
\item{The adoption patterns have been highly uniform across popular domains; most domains tend to follow similar upgrading paths used also by other domains (\ref{rq: uniform}).}
\item{Only the prevalence of adoption (\ref{rq: prevalence}) varies between the short-run history (2016 -- early 2017) and long-run evolution (2012 -- early 2017). Namely, the longer the period observed, the more common has adoption been.}
\end{itemize}

These findings provide also some material for contemplating about the current release engineering strategy of the PHP project. For the developers of the programming language, a pressing question relates to the means by which the currently slow adoption of PHP~7 could be boosted in the future. One option to consider might be a Firefox-style rapid release strategy, which has been suspected to increase user adoption compared to a traditional release strategy \cite{daCosta16}. Because adoption has generally been infrequent among popular web sites, it might be possible to also debate whether the current release schedule is actually already too rapid for users and stakeholders. Although downgrading is rare and the adoption patterns are generally uniform, a further interesting question relates to the reasons why some domains downgrade, and whether there is anything that could be done to help outlying domains following chaotic release adoption paths.

\subsection{Threats to Validity}

Threats to validity can be enumerated by using the conventional threefold classification of construct validity, external validity, and internal validity. Although there exists no uniformly agreed definitions~\cite{Siegmund15}, for the purposes of this paper, these three validity concepts can be equated to questions related to generalizability (how results generalize to a different context or population), operationalization (how well a quantification matches a theoretical concept), and systematic computational errors (how well different biases are eliminated), respectively.

\subsubsection{External Validity}

Generalizability questions are always present when the theoretical population is the whole world wide web \cite{Ainsworth13}, including the so-called ``deep web'' not indexed by standard search engines. Even though generalizability toward such a population is practically impossible even for companies such as Google, it is possible to narrow the target of generalization toward a sub-population of the most popular PHP-powered domains. In this regard, HTTP Archive uses Alexa's popularity list, which is commonly perceived as a good choice for  seeding of large-scale web crawling~\text{\cite{Barford14, Park14, Tappenden09}}. While external validity is presumably not threatened in this regard, (a) the results reported are likely specific to popular web sites. When considering further applications, such as those motivated by security questions~\cite{Medeiros16}, it is likely that more interesting cases are located in the fringes of the world wide web. Given that prevalence of PHP release adoption was observed to be at a modest level in a sample of popular sites, it is more than likely that even lower levels of adoption could be observed in a sample covering WordPress deployments, for instance. A common limitation \cite{He13} is also present: (b) the results apply only to domains using PHP for serving pages via plain HTTP, excluding sites using HTTPS.

\subsubsection{Construct Validity}

A notable threat to construct validity stems from the identification of PHP deployments via a regular expression from HTTP response headers (see Section~\ref{subsec: pre-processing}). This coarse identification technique is likely to include both false positives (popular domains incorrectly identified as running with PHP) and false negatives (the missing of popular PHP-powered domains). To evaluate the severity of this limitation, at minimum, parallel identification should be attempted from the actual web page content (cf.~\cite{WangNappa14}). Because the primary identification requirement relates to the version of a PHP backend used for serving a particular web content, robust identification is likely challenging also from web page contents, however. Further research is therefore required to continue the work on identifying and fingerprinting PHP applications~\cite{Kozina09}, including the PHP interpreter itself.

\subsubsection{Internal Validity}

The potential presence of systematic biases is best evaluated against the classical DTMC assumptions that were imposed for the statistical computation (see Section~\ref{subsec: dtmcs in brief}). There are three notable concerns about these assumptions. First, the assumption in Eq.~\eqref{eq: Markov property} implies that regardless whether a state change is due to security updates, reliability improvements, or new features, it is always the currently deployed version that defines the reference point for the change, regardless whether the decision to change versions is made by a human or a package manager. While the assumption seems sensible from a release engineering viewpoint, it is easily questioned from a software evolution perspective \cite{WongCai11, Ruohonen15JSEP}. If history matters also for PHP release adoption patterns, it would seem reasonable to recommend that further research should focus on higher-order Markov chains that have a memory~\text{\cite{WongCai11, Ching08, Singer14}}. The second concern relates to the assumption of independence between domains. Given that a substantial amount of contemporary web sites require connections to two or more servers~\text{\cite{Wambach16, Newton13}}, PHP deployments may be uniformly managed and upgraded in a cloud computing service or other large deployment farm. Consequently, a PHP version sequence of a domain might be affected by a sequence of another domain. Conditional Markov chains \cite{Ching08, Goutte14} may provide a useful tool for evaluating the potential severity of this cross-domain dependence assumption. 

The third notable threat to internal validity relates to the PHP version sequences observed, which mandate making an addition assumption about the transition probabilities in \eqref{eq: MLE}. Consequently, by definition \cite{Seber08}, the transition probability matrices computed are not stochastic matrices, that is, the row sums of these matrices do not necessarily equal one. Although this unavoidable limitation does not affect the results reported as such, it does affect additional computations involving eigenvalues \cite{Hill04}, and particularly the stationary distributions toward which all irreducible, aperiodic, and positive recurrent Markov chains converge (for the mathematical background see~\cite{Privault13}). This point should be kept in mind when considering further DTMC applications in the release engineering and software evolution contexts. Such applications are also a good way to point out a couple of new research directions.

\subsection{Further Work}

The primary purpose of this paper was to examine the usefulness of DTMC modeling for systematic tracing of web deployments in order to establish automated continuous feedback channel for server-side programming language developers and stakeholders. The paper fulfilled this goal: DTMCs are useful also in the release engineering context. For pursuing DTMC analysis further, a worthwhile goal would be to translate some of the concepts used in other disciplines to the language of release engineering and software evolution. For instance, simple DTMC metrics have been used to proxy such concepts as colonization, disturbance, and replacement~\cite{Hill04}. With some theoretical and terminological alterations, such metrics and concepts could be adopted for pursuing DTMC modeling further in the release engineering context. While these concepts and metrics are directly applicable to traditional DTMCs, another prolific path forward involves altering the basic assumptions surrounding discrete time-homogeneous Markov chains. Conditional and higher-order chains are good examples in this regard. For continuous tracing of PHP deployments, continuous Markov chains (as opposed to discrete-time chains) seem prolific to consider in further research. For instance, different time-delay models~\cite{Wang12} could be adopted for studying the time delays between successive state changes. The question about time delays is also fundamental in the release engineering context because the empirical transition probabilities depend on the sampling frequency used.

\section*{Acknowledgements}

The authors gratefully acknowledge Tekes -- the Finnish Funding Agency for Innovation, DIMECC Oy, and the Cyber Trust research program for their support.

%\clearpage
%\newpage
\balance
\bibliographystyle{IEEEtran}
%\bibliography{phpmarkov}

\begin{thebibliography}{10}
\providecommand{\url}[1]{#1}
\csname url@samestyle\endcsname
\providecommand{\newblock}{\relax}
\providecommand{\bibinfo}[2]{#2}
\providecommand{\BIBentrySTDinterwordspacing}{\spaceskip=0pt\relax}
\providecommand{\BIBentryALTinterwordstretchfactor}{4}
\providecommand{\BIBentryALTinterwordspacing}{\spaceskip=\fontdimen2\font plus
\BIBentryALTinterwordstretchfactor\fontdimen3\font minus
  \fontdimen4\font\relax}
\providecommand{\BIBforeignlanguage}[2]{{%
\expandafter\ifx\csname l@#1\endcsname\relax
\typeout{** WARNING: IEEEtran.bst: No hyphenation pattern has been}%
\typeout{** loaded for the language `#1'. Using the pattern for}%
\typeout{** the default language instead.}%
\else
\language=\csname l@#1\endcsname
\fi
#2}}
\providecommand{\BIBdecl}{\relax}
\BIBdecl

\bibitem{Favre05}
J.-M. Favre, ``{L}anguages {E}volve {T}oo! {C}hanging the {S}oftware {T}ime
  {S}cale,'' in \emph{Proceedings of the Eighth International Workshop on
  Principles of Software Evolution (IWPSE 2005)}.\hskip 1em plus 0.5em minus
  0.4em\relax Lisbon: IEEE, 2005, pp. 33--42.

\bibitem{Meyerovich12}
L.~A. Meyerovich and A.~S. Rabkin, ``{S}ocio-{PLT}: {P}rinciples for
  {P}rogramming {L}anguage {A}doption,'' in \emph{Proceedings of the ACM
  International Symposium on New Ideas, New Paradigms, and Reflections on
  Programming and Software}.\hskip 1em plus 0.5em minus 0.4em\relax Tucson:
  ACM, 2012, pp. 39--54.

\bibitem{Cheung80}
R.~C. Cheung, ``{A} {U}ser-{O}riented {S}oftware {R}eliability {M}odel,''
  \emph{IEEE Transactions on Software Engineering}, vol. SE-6, no.~2, pp.
  118--125, 1980.

\bibitem{Wang12}
W.~Wang, ``{A}n {O}verview of the {R}ecent {A}dvances in {D}elay-{T}ime-{B}ased
  {M}aintenance {M}odeling,'' \emph{Reliability Engineering and System Safety},
  vol. 106, pp. 165--178, 2012.

\bibitem{WongCai11}
S.~Wong and Y.~Cai, ``{G}eneralizing {E}volutionary {C}oupling with
  {S}tochastic {D}ependencies,'' in \emph{Proceedings of the 26th IEEE/ACM
  International Conference on Automated Software Engineering (ASE 2011)}.\hskip
  1em plus 0.5em minus 0.4em\relax Lawrence: IEEE, 2011, pp. 293--302.

\bibitem{Raemaekers14}
S.~Raemaekers, A.~{van Deursen}, and J.~Visser, ``{S}emantic {V}ersioning
  versus {B}reaking {C}hanges: {A} {S}tudy of the {M}aven {R}epository,'' in
  \emph{Proceedings of the IEEE 14th International Working Conference on Source
  Code Analysis and Manipulation (SCAM 2014)}.\hskip 1em plus 0.5em minus
  0.4em\relax Victoria: IEEE, 2014, pp. 215--224.

\bibitem{Kula15}
R.~G. Kula, D.~M. German, T.~Ishio, and K.~Inoue, ``{T}rusting a {L}ibrary: {A}
  {S}tudy of the {L}atency to {A}dopt the {L}atest {M}aven {R}elease,'' in
  \emph{Proceedings of the IEEE 22nd International Conference on Software
  Analysis, Evolution, and Reengineering (SANER 2015)}, Montreal, 2015, pp.
  520--524.

\bibitem{Karvonen17}
T.~Karvonen, W.~Behutiye, M.~Oivo, and P.~Kuvaja, ``{S}ystematic {L}iterature
  {R}eview on the {I}mpacts of {A}gile {R}elease {E}ngineering {P}ractices,''
  \emph{Information and Software Technology}, vol.~86, pp. 87--100, 2017.

\bibitem{Mantyla13}
M.~V. M\"antyl\"a, F.~Khomh, B.~Adams, E.~Engstr\"om, and K.~Petersen, ``{O}n
  {R}apid {R}eleases and {S}oftware {T}esting,'' in \emph{Proceedings of the
  IEEE International Conference on Software Maintenance (ICSME 2013)}.\hskip
  1em plus 0.5em minus 0.4em\relax Madrid: IEEE, 2013, pp. 20--29.

\bibitem{daCosta16}
D.~A. {da Costa}, S.~McIntosh, U.~Kulesza, and A.~E. Hassan, ``{T}he {I}mpact
  of {S}witching to a {R}apid {R}elease {C}ycle on the {I}ntegration {D}elay of
  {A}ddressed {I}ssues: {A}n {E}mpirical {S}tudy of the {M}ozilla {F}irefox
  {P}roject,'' in \emph{Proceedings of the 13th International Conference on
  Mining Software Repositories (MSR 2016)}.\hskip 1em plus 0.5em minus
  0.4em\relax Austin: ACM, 2016, pp. 374--385.

\bibitem{Adams16}
B.~Adams and S.~McIntosh, ``{M}odern {R}elease {E}ngineering in a {N}utshell --
  {W}hy {R}esearchers {S}hould {C}are,'' in \emph{Proceedings IEEE 23rd
  International Conference on Software Analysis, Evolution, and Reengineering
  (SANER 2016)}.\hskip 1em plus 0.5em minus 0.4em\relax Osaka: IEEE, 2016, pp.
  78--90.

\bibitem{Baysal12}
O.~Baysal, R.~Holmes, and M.~W. Godfrey, ``{M}ining {U}sage {D}ata and
  {D}evelopment {A}rtifacts,'' in \emph{Proceedings of the 9th IEEE Working
  Conference on Mining Software Repositories (MSR 2012)}.\hskip 1em plus 0.5em
  minus 0.4em\relax Zurich: IEEE, 2012, pp. 98--107.

\bibitem{Ruohonen16WIMS}
J.~Ruohonen, S.~Hyrynsalmi, and V.~Lepp\"anen, ``{E}xploring the {U}se of
  {D}eprecated {PHP} {R}eleases in the {W}ild {I}nternet: {S}till a {LAMP}
  {I}ssue?'' in \emph{Proceedings of the 6th International Conference on Web
  Intelligence, Mining and Semantics (WIMS 2016)}.\hskip 1em plus 0.5em minus
  0.4em\relax N\^imes: ACM, 2016, pp. \text{26:1--26:12}.

\bibitem{Amanatidis16}
T.~Amanatidis and A.~Chatzigeorgiou, ``{S}tudying the {E}volution of {PHP}
  {W}eb {A}pplications,'' \emph{Information and Software Technology}, vol.~72,
  pp. 48--67, 2016.

\bibitem{Hills15}
M.~Hills, ``{E}volution of {D}ynamic {F}eature {U}sage in {PHP},'' in
  \emph{Proceedings of the IEEE 22nd International Conference on Software
  Analysis, Evolution, and Reengineering (SANER 2015)}.\hskip 1em plus 0.5em
  minus 0.4em\relax Montreal: IEEE, 2015, pp. 525--529.

\bibitem{Hills13}
M.~Hills, P.~Klint, and J.~Vinju, ``{A}n {E}mpirical {S}tudy of {PHP} {F}eature
  {U}sage: {A} {S}tatic {A}nalysis {P}erspective,'' in \emph{Proceedings of the
  International Symposium on Software Testing and Analysis (ISSTA 2013)}.\hskip
  1em plus 0.5em minus 0.4em\relax Lugano: ACM, 2013, pp. 325--335.

\bibitem{Medeiros16}
I.~Medeiros, N.~Neves, and M.~Correia, ``{D}etecting and {R}emoving {W}eb
  {A}pplication {V}ulnerabilities with {S}tatic {A}nalysis and {D}ata
  {M}ining,'' \emph{IEEE Transactions on Reliability}, vol.~65, no.~1, pp.
  54--69, 2016.

\bibitem{Fitzgerald15}
B.~Fitzgerald and K.~Stol, ``{C}ontinuous {S}oftware {E}ngineering:
  {A}~{R}oadmap and {A}genda,'' \emph{Journal of Systems and Software}, vol.
  123, pp. \text{176--189}, 2015.

\bibitem{PangHindle16}
C.~Pang and A.~Hindle, ``{C}ontinuous {M}aintenance,'' in \emph{Proceedings of
  the IEEE International Conference on Software Maintenance and Evolution
  (ICSME 2016)}.\hskip 1em plus 0.5em minus 0.4em\relax Raleigh: IEEE, 2016,
  pp. 458--462.

\bibitem{Dyck15}
A.~Dyck, R.~Penners, and H.~Lichter, ``{T}owards {D}efinitions for {R}elease
  {E}ngineering and {D}ev{O}ps,'' in \emph{Proceedings of the Third
  International Workshop on Release Engineering (RELENG 2015)}.\hskip 1em plus
  0.5em minus 0.4em\relax Florence: IEEE, 2015, pp. 3--3.

\bibitem{Leppanen15}
M.~Lepp\"anen, S.~M\"akinen, M.~Pagels, V.-P. Eloranta, J.~Itkonen, M.~V.
  M\"antyl\"a, and T.~M\"annist\"o, ``{T}he {H}ighways and {C}ountry {R}oads to
  {C}ontinuous {D}eployment,'' \emph{IEEE Software}, vol.~32, no.~2, pp.
  64--72, 2015.

\bibitem{deOliveira16}
R.~P. {de Oliveira}, A.~R. Santos, E.~S. {de Almeida}, and G.~S.
  da~Silva~Gomes, ``{E}valuating {L}ehman's {L}aws of {S}oftware {E}volution
  {W}ithin {S}oftware {P}roduct {L}ines {I}ndustrial {P}rojects,''
  \emph{Journal of Systems and Software}, vol. 131, pp. 347--365, 2016.

\bibitem{Ruohonen15JSEP}
J.~Ruohonen, S.~Hyrynsalmi, and V.~Lepp\"anen, ``{T}ime {S}eries {T}rends in
  {S}oftware {E}volution,'' \emph{Journal of Software: Evolution and Process},
  vol.~27, no.~2, pp. 990--1015, 2015.

\bibitem{Gotel94}
O.~C.~Z. Gotel and A.~C.~W. Finkelstein, ``{A}n {A}nalysis of the
  {R}equirements {T}raceability {P}roblem,'' in \emph{Proceedings of IEEE
  International Conference on Requirements Engineering (ICRE 1994)}.\hskip 1em
  plus 0.5em minus 0.4em\relax IEEE, 1994, pp. 94--101.

\bibitem{Sturgeon14}
P.~Sturgeon, ``{T}he {N}everending {M}uppet {D}ebate of {PHP 6} v {PHP 7},''
  2014, {A}vailable online in March 2017:
  \url{https://philsturgeon.uk/php/2014/07/23/neverending-muppet-debate-of-php-6-v-php-7/}.

\bibitem{PHP10}
{The PHP Project}, ``{R}equest for {C}omments: {R}elease {P}rocess,'' 2010,
  {A}vailable online in March 2017:
  \url{https://wiki.php.net/rfc/releaseprocess}.

\bibitem{PHP17b}
------, ``{S}upported {V}ersions,'' 2017, {A}vailable online in March 2017:
  \url{http://php.net/supported-versions.php}.

\bibitem{PHP17a}
------, ``{U}nsupported {H}istorical {R}eleases,'' 2017, {A}vailable online in
  March 2017: \url{https://secure.php.net/releases/}.

\bibitem{Mantyla11}
M.~V. M\"antyl\"a and J.~Vanhanen, ``{S}oftware {D}eployment {A}ctivities and
  {C}hallenges -- {A} {C}ase {S}tudy of {F}our {S}oftware {P}roduct
  {C}ompanies,'' in \emph{Proceedings of the 15th European Conference on
  Software Maintenance and Reengineering (CSMR 2011)}.\hskip 1em plus 0.5em
  minus 0.4em\relax Oldenburg: IEEE, 2011, pp. \text{131--140}.

\bibitem{Singer14}
P.~Singer, D.~Helic, B.~Taraghi, and M.~Strohmaier, ``{D}etecting {M}emory and
  {S}tructure in {H}uman {N}avigation {P}atterns {U}sing {M}arkov {C}hain
  {M}odels of {V}arying {O}rder,'' \emph{PLOS ONE}, vol.~9, no.~7, p. e102070,
  2014.

\bibitem{Hill04}
M.~F. Hill, J.~D. Witman, and H.~Caswell, ``{M}arkov {C}hain {A}nalysis of
  {S}uccession in a {R}ocky {S}ubtidal {C}ommunity,'' \emph{The American
  Naturalist}, vol. 164, no.~2, pp. E46--E61, 2004.

\bibitem{markovchain}
G.~A. Spedicato, ``{markovchain}: {D}iscrete {T}ime {M}arkov {C}hains {M}ade
  {E}asy,'' 2016, {R} package version 0.6, available online in March 2017:
  \url{https://cran.r-project.org/web/packages/markovchain/index.html}.

\bibitem{HTTPArchive17a}
{HTTP Archive}, ``{D}ownloads,'' 2017, {A}vailable online in March 2017:
  \url{http://httparchive.org/downloads.php}.

\bibitem{Wambach16}
T.~Wambach and K.~Br\"aunlich, ``{T}he {E}volution of {T}hird-{P}arty {W}eb
  {T}racking,'' in \emph{Proceedings of the International Conference on
  Information Systems Security and Privacy (ICISSP 2016)}, O.~Camp, S.~Furnell,
  and P.~Mori, Eds.\hskip 1em plus 0.5em minus 0.4em\relax Rome: Springer,
  2016.

\bibitem{Ainsworth13}
S.~G. Ainsworth and M.~L. Nelson, ``{E}valuating {S}liding and {S}ticky
  {T}arget {P}olicies by {M}easuring {T}emporal {D}rift in {A}cyclic {W}alks
  {T}hrough a {W}eb {A}rchive,'' in \emph{Proceedings of the 13th ACM/IEEE-CS
  Joint Conference on Digital Libraries (JCDL 2013)}.\hskip 1em plus 0.5em
  minus 0.4em\relax Indianapolis: ACM, 2013, pp. 39--48.

\bibitem{HTTPArchive17b}
{HTTP Archive}, ``{FAQ},'' 2017, {A}vailable online in March 2017:
  \url{http://httparchive.org/about.php#faq}.

\bibitem{Calzarossa14}
M.~C. Calzarossa and L.~Massari, ``{A}nalysis of {H}eader {U}sage {P}atterns of
  {HTTP} {R}equest {M}essages,'' in \emph{Proceedings of the 2014 IEEE Intl
  Conf on High Performance Computing and Communications, 2014 IEEE 6th Intl
  Symp on Cyberspace Safety and Security, 2014 IEEE 11th Intl Conf on Embedded
  Software and Syst (HPCC, CSS, ICESS)}.\hskip 1em plus 0.5em minus 0.4em\relax
  Paris: IEEE, 2014, pp. 847--853.

\bibitem{Pham16}
K.~Pham, A.~Santos, and J.~Freire, ``{U}nderstanding {W}ebsite {B}ehavior
  {B}ased on {U}ser {A}gent,'' in \emph{Proceedings of the 39th International
  ACM SIGIR Conference on Research and Development in Information Retrieval
  (SIGIR 2016)}.\hskip 1em plus 0.5em minus 0.4em\relax Pisa: ACM, 2016, pp.
  1053--1056.

\bibitem{WangNappa14}
L.~Wang, A.~Nappa, J.~Caballero, T.~Ristenpart, and A.~Akella, ``{W}ho{W}as:
  {A} {P}latform for {M}easuring {W}eb {D}eployments on {IaaS} {C}louds,'' in
  \emph{Proceedings of the 2014 Conference on Internet Measurement Conference
  (IMC 2014)}.\hskip 1em plus 0.5em minus 0.4em\relax Vancouver: ACM, 2014, pp.
  101--114.

\bibitem{Siegmund15}
J.~Siegmund, N.~Siegmund, and S.~Apel, ``{V}iews on {I}nternal and {E}xternal
  {V}alidity in {E}mpirical {S}oftware {E}ngineering,'' in \emph{Proceedings of
  the IEEE/ACM 37th IEEE International Conference on Software Engineering (ICSE
  2015)}.\hskip 1em plus 0.5em minus 0.4em\relax Florence: IEEE, 2015, pp.
  9--19.

\bibitem{Barford14}
P.~Barford, I.~Canadi, D.~Krushevskaja, Q.~Ma, and S.~Muthukrishnan,
  ``{A}dscape: {H}arvesting and {A}nalyzing {O}nline {D}isplay {A}ds,'' in
  \emph{International Conference on World Wide Web (WWW 2014)}.\hskip 1em plus
  0.5em minus 0.4em\relax Seoul: ACM, 2014, pp. 597--608.

\bibitem{Park14}
Y.~J. Park, ``{A} {B}roken {S}ystem of {S}elf-{R}egulation of {P}rivacy
  {O}nline? {S}urveillance, {C}ontrol, and {L}imits of {U}ser {F}eatures in
  {U.S.} {W}ebsites,'' \emph{Policy \& Internet}, vol.~6, no.~4, pp. 360--376,
  2014.

\bibitem{Tappenden09}
A.~F. Tappenden and J.~Miller, ``{C}ookies: {A} {D}eployment {S}tudy and the
  {T}esting {I}mplications,'' \emph{ACM Transactions on the Web}, vol.~3,
  no.~3, pp. 9:1 -- 9:49, 2009.

\bibitem{He13}
K.~He, A.~Fisher, L.~Wang, A.~Gember, A.~Akella, and T.~Ristenpart, ``{N}ext
  {S}top, the {C}loud: {U}nderstanding {M}odern {W}eb {S}ervice {D}eployment in
  {EC2} and {A}zure,'' in \emph{Proceedings of the 2013 Conference on Internet
  Measurement Conference (IMC 2013)}.\hskip 1em plus 0.5em minus 0.4em\relax
  Barcelona: ACM, 2013, pp. \text{177--190}.

\bibitem{Kozina09}
M.~Kozina, M.~Golub, and S.~Gro\v{s}, ``{A} {M}ethod for {I}dentifying {W}eb
  {A}pplications,'' \emph{International Journal of Information Security},
  vol.~8, no.~6, pp. \text{455--467}, 2009.

\bibitem{Ching08}
W.-K. Ching, M.~K. Ng, and E.~S. Fung, ``{H}igher-{O}rder {M}ultivariate
  {M}arkov {C}hains and {T}heir {A}pplicatons,'' \emph{Linear Algebra and Its
  Applications}, vol. 428, no. 2--3, pp. 492--507, 2008.

\bibitem{Newton13}
B.~Newton, K.~Jeffay, and J.~Aikat, ``{T}he {C}ontinued {E}volution of {W}eb
  {T}raffic,'' in \emph{Proceedings of the IEEE 21st International Symposium on
  Modelling, Analysis and Simulation of Computer and Telecommunication Systems
  (MASCOTS 2013)}.\hskip 1em plus 0.5em minus 0.4em\relax IEEE, 2013, pp.
  80--89.

\bibitem{Goutte14}
S.~Goutte, ``{C}onditional {M}arkov {R}egime {S}witching {M}odel {A}pplied to
  {E}conomic {M}odelling,'' \emph{Economic Modelling}, vol.~38, pp. 258--269,
  2014.

\bibitem{Seber08}
G.~A.~F. Seber, \emph{{A} {M}atrix {H}andbook for {S}tatisticians}.\hskip 1em
  plus 0.5em minus 0.4em\relax New Jersey: John Wiley \& Sons, 2008.

\bibitem{Privault13}
N.~Privault, \emph{{U}nderstanding {M}arkov {C}hains: {E}xamples and
  {A}pplications}.\hskip 1em plus 0.5em minus 0.4em\relax Heidelberg: Springer,
  2013.

\end{thebibliography}

% Generated by IEEEtran.bst, version: 1.13 (2008/09/30)

\end{document}